\documentstyle[dina4,12pt]{article}
\newcommand{\eq}{\label}

\begin{document}
\title{Antiproton Production in p-Nucleus and Nucleus-Nucleus Collisions
 within a Relativistic Transport Approach
    \protect \footnote[2]{Supported by BMFT, GSI Darmstadt and KFA J\"ulich.} }
\author{Stefan Teis, Wolfgang Cassing, Tomoyuki Maruyama \\
                and Ulrich Mosel \\
        Institut f\"{u}r Theoretische Physik, Universit\"{a}t Giessen \\
        D-35392 Giessen, Germany}
\maketitle
\begin{abstract}
The production of antiprotons in proton-nucleus
 and nucleus-nu\-cle\-us reactions is
 calculated within the relativistic BUU approach employing proper
selfenergies for the baryons and antiprotons and treating the $\bar{p}$
annihilation nonperturbatively. The differential cross section for the
antiprotons is found to be very sensitive to the $\bar{p}$ selfenergy adopted.
A detailed comparison with the available experimental data
for p-nucleus and nucleus-nucleus reactions  shows
that the antiproton feels a moderately attractive mean-field at normal
nuclear matter density $\rho_0$ which is in line with a dispersive
potential extracted from the free annihilation cross section.

\end{abstract}

\vspace{2cm}

The production of particles at energies below the free nucleon-nucleon
threshold ('subthreshold production') constitutes one of
the most promising sources of information about
the properties of nuclear matter at high densities since the particles are
produced predominantly during the compressed stage at high density
\cite{cassing,mosel,Aich}.
Antiproton production at energies of a few GeV/u is the most
extreme subthreshold production process and has been
observed in proton-nucleus collisions already more than 20 years ago
\cite{chamberlain,elioff,dorfan}.
Experiments at the JINR \cite{JINR} and at the BEVALAC
\cite{BEVALAC1,BEVALAC2} have provided, furthermore, first measurements of
subthreshold antiproton production in nucleus-nucleus collisions. Since then
the problem was taken up again at KEK \cite{KEK} and GSI \cite{GSI}
with new detector setups.    Various
descriptions for these data have been proposed.
Based on thermal models it has been suggested that the antiproton yield
contains large contributions from $\Delta N \rightarrow \bar{p} +X$,
$\Delta \Delta \rightarrow \bar{p} + X$  and $\rho \rho \rightarrow \bar{p}N$
production mechanisms \cite{koch,ko1,ko2}.
Other models have attempted to explain these data in terms of multiparticle
interactions \cite{danielewicz}.

In a first chance nucleon-nucleon
collision model (assuming high momentum tails consistent with data on backward
proton scattering) Shor et al. \cite{shor} succeeded in reproducing the
 antiproton yield  for the proton-nucleus case, however,
 underestimated the yield by more than 3 orders of magnitude for
nucleus-nucleus collisions. This problem was partly resolved by Batko et
al. \cite{Batko} who performed the first nonequilibrium $\bar{p}$-production
study on the basis of the VUU transport equation. Within this approach it
became possible to describe simultaneously the $\bar{p}$-data from p + A
and A + A reactions; however, the yield was still underestimated
 when including
the strong $\bar{p}$ annihilation. Nevertheless, it became clear, that in
A + A reactions the dominant production channel proceeds via an
intermediate nucleon resonance which allows to store a sizeable amount of
energy that can be used in a subsequent collision for the production of a
$ p \bar{p}$ pair. Later on, these results were also confirmed by Huang et al.
\cite{Faessl} within the QMD model; the authors achieved a reproduction
 of nucleus-nucleus data only when neglecting the annihilation channel.

These results have led to the suggestion that the quasi-particle
properties of the nucleons might be important for the $\bar{p}$ production
process which become more significant with increasing nuclear density.
Schaffner et al. \cite{Mis} found in a thermal relativistic model, assuming
kinetic and chemical equilibrium, that the $\bar{p}$-abundancy might be
dramatically enhanced when assuming the antiproton selfenergy to be
given by charge conjugation of the nucleon selfenergy. This leads to
strong attractive vector selfenergies of the antiproton. However, the
above concept lacks unitarity between the real and the imaginary part
of the $\bar{p}$-selfenergy and thus remains questionable. Besides this,
even in the $\sigma - \omega$ model the Fock terms lead to a suppression
of the attractive $\bar{p}$-field \cite{Soutome}, such
 that the production threshold
is shifted up in energy again as compared to the simple model involving
 charge conjugation. Furthermore,
the assumption of thermal and especially chemical equilibrium most
likely is not fullfilled e.g. in Si + Si collisions around 2 GeV/u
\cite{Lang}.

First fully relativistic transport calculations for antiproton
production including $\bar{p}$ annihilation as well as the change of the
quasi-particle properties in the medium have been reported in \cite{Cass92}.
There it was found that according to the reduced nucleon mass in the
medium the threshold for $\bar{p}$-production is shifted to lower energy
and the antiproton cross section prior to annihilation becomes enhanced
for Si + Si at 2.1 GeV/u by roughly 2 orders of magnitude as compared to a
relativistic cascade calculation where no in-medium effects are incorporated.
The enhancement of the antiproton yield then was dramatically reduced again
when including the strong annihilation channel which, however, lead -
accidentally, as noted in \cite{Cass92} - to a reasonable reproduction
of the $\bar{p}$ data for Si + Si at 1.65 and 2.1 GeV/u. A final answer on
the antiproton problem especially with respect to the $\bar{p}$ selfenergy
in the medium could not be given in \cite{Cass92} since the explicit
momentum dependence of the nucleons scalar and vector selfenergies had not
been accounted for.

In this letter we therefore analyze the production of antiprotons within
the framework of the relativistic transport theory (RBUU) that calculates
$\bar{p}$ production perturbatively as in \cite{Batko,Cass92} from the
channels $ N N \rightarrow N N p \bar{p}, \Delta N \rightarrow N N p \bar{p}$
as well as $\Delta \Delta \rightarrow N N p \bar{p}$. The quasi-particle
properties, i.e. the nucleon selfenergies $U_s({\bf p}, \rho) ,
U_{\mu}({\bf p}, \rho)$, are taken from \cite{Tomo,KLW1,KLW2}
and are in line with the results of Dirac-Brueckner calculations
\cite{Mal} whereas the antiproton selfenergies are described on the
basis of the $\sigma - \omega$ model \cite{Serot} with free coupling
parameters $g_s^{\bar{p}}$ and $g_{v}^{\bar{p}}$. Antiprotons are
propagated explicitly in the respective time dependent potentials and their
annihilation is calculated nonperturbatively by means of individual rate
equations. A comparison with the most recent data from KEK and GSI on
$\bar{p}$-production will allow to approximately determine
the free parameters $g_s^{\bar{p}}$ and $g_{v}^{\bar{p}}$ and
 provide first information on the antiproton potential in the medium.
We finally compare the numerical results for the Schr\"odinger equivalent
antiproton potential with a dispersive potential as evaluated from the
annihilation rate and the charge conjugate potential as expected from
simple relativistic mean-field theory.

Since the covariant BUU approach has been extensively discussed in the
reviews \cite{BUU1,Blaettel} we only recall the basic equations and the
corresponding quasi-particle properties that are required for a proper
understanding of the results reported in this study.
The relativistic BUU (RBUU) equation with momentum-dependent mean-fields or
selfenergies is given by
\begin{equation}
\{ [ \Pi_\mu - \Pi_\nu ( {\partial}^{p}_{\mu} U^{\nu} ) +
M^* ( {\partial}^{p}_{\mu} U_s ) ]  \partial^\mu_x  +
 [ - \Pi_\nu ( {\partial}^{x}_{\mu} U^{\nu} ) +
M^* ( {\partial}^{x}_{\mu} U_s ) ] \partial^\mu_p \} f(x,p) = I_{coll} ,
\eq{090}
\end{equation}
where $f(x,p)$ is the Lorentz covariant phase-space distribution function,
$I_{coll}$ is a collision term given in Ref. \cite{BUU1}, and $U_s$ and
$U_\mu$ are the scalar- and the vector- mean-fields.
The effective mass $M^*$ and the kinetic momentum $\Pi_\mu$ are defined in
terms of the fields by
\begin{eqnarray}
\Pi_\mu (x,p) & = & p_\mu - U_\mu (x,p)
\\
M^* (x,p) & = & M + U_s (x,p) ,
\eq{e200}
\end{eqnarray}
while the quasi-particle mass-shell constraint is obtained as
\begin{equation}
V(x,p) f(x,p) = 0
\eq{e300}
\end{equation}
with the pseudo potential
\begin{equation}
V(x,p) \equiv \frac{1}{2} \{ \Pi^2 (x,p) - M^{*2} (x,p) \} .
\eq{e400}
\end{equation}
The above equation implies that the phase-space distribution function $f(x,p)$
is nonvanishing only on the hypersurface in phase-space defined by
 $V(x,p) = 0$.

In order to implement proper selfenergies for the nucleons in the $\bar{p}$
production processes we follow ref.  \cite{KLW1} and separate
 the mean-fields into a local part
and an explicit momentum-dependent part,
i.e.
\begin{eqnarray}
U_s(x,p) & = & U_s^H (x) + U_s^{MD} (x,p)
\nonumber \\
U_{\mu}(x,p) & = & U_{\mu}^H (x) + U_{\mu}^{MD} (x,p)  .
\eq{e500}
\end{eqnarray}
where the local mean-fields are determined by the usual Hartree equation:
\begin{eqnarray}
U_s^{H} (x) & = & - g_s \sigma_H (x)
\nonumber \\
U_\mu^{H} (x) & = & g_v \omega^H_\mu (x)
\eq{e600}
\end{eqnarray}
with
\begin{eqnarray}
m_s^{2} \sigma_H (x) + B_s \sigma_H^2 (x) + C_s \sigma_H^3 (x)
= g_s \rho_s (x)
\nonumber \\
m_s^{2} \omega_\mu^H (x) = g_v j_\mu (x) .
\eq{e700}
\end{eqnarray}
In the above equations
the scalar density $\rho_s (x)$ and the current $j_\mu (x)$ are given
in terms of the phase-space distribution function by
\begin{eqnarray}
\rho_s (x) = \frac{4}{(2 \pi)^3} \int d^4 p \  M^* (x,p) f(x,p)
\nonumber \\
j_\mu (x) = \frac{4}{(2 \pi)^3} \int d^4 p \ \Pi_\mu (x,p) f(x,p).
\eq{e800}
\end{eqnarray}

The free parameters of the above expressions are fixed to reproduce the
saturation properties of nuclear matter, the empirical proton-nucleus
optical potential as well as the density dependence of $U_s$ and
$U_{\mu}$ from Dirac-Brueckner theory \cite{Mal}. Explicit values for
these parameters are given in \cite{Tomo,KLW1,KLW2}.
 The actual results for $U_s(p)$
and the zero'th component of the vector field $U_0(p)$ are displayed in
Fig.1 for $\rho_0 (\approx 0.17 fm^{-3}), 2 \rho_0,$ and $ 3 \rho_0$.

As mentioned before, the phase-space distribution function for the
antiprotons $f_{\bar{p}}(x,p)$ is assumed to follow an equation of motion as
in  (1), however, with scalar and vector potentials of different strength, i.e.
\begin{eqnarray}
U_s^{\bar{p}} (x) & = & - g_s^{\bar{p}} \sigma_H (x)
\nonumber \\
U_\mu^{\bar{p}} (x) & = & g_v^{\bar{p}} \omega^H_\mu (x)
\eq{e660}
\end{eqnarray}
where $g_s^{\bar{p}}$ and $g_v^{\bar{p}}$ are treated as independent parameters
and are fixed in comparison to the experimental data (see below). The
collision term (r.h.s. of eq. (1)) for the antiproton phase-space
distribution besides elastic scattering also includes a direct coupling
to the nucleons which describes the $\bar{p}$-annihilation channel
(cf. discussion below).

The antiproton invariant differential multiplicity is obtained by summing
incoherently the elementary antiproton multiplicities over all collisions
and integrating over the residual degrees of freedom \cite{cassing}.
If we consider that the antiprotons are produced via processes of the type
$BB \rightarrow \bar{p} p + NN \equiv \bar{p} + 3 + 4 + 5$ ($B$ stands for
either nucleon or   $\Delta$)  we can write the antiproton invariant
multiplicity as \cite{Batko}
\begin{eqnarray}
\lefteqn{
    E_{\bar{p}} \frac{d^3N(b)}{d^3p_{\bar{p}}}  }    \nonumber \\
& & = \sum_{BB coll} \int d^3p'_3 d^3p'_4 d^3p'_5  \,
    \frac{1}{\sigma_{BB}(\sqrt{s})}
    E'_{\bar{p}} \frac{d^{12} \sigma_{BB \rightarrow \bar{p}}(\sqrt{s})}
    {d^3p'_{\bar{p}} d^3p'_3 d^3p'_4 d^3p'_5}          \nonumber \\
& & \left[ 1 - f(\vec{r}, \vec{p'}_3; t) \right]
    \left[ 1 - f(\vec{r}, \vec{p'}_4; t) \right]
    \left[ 1 - f(\vec{r}, \vec{p'}_5; t) \right]
                                            \label{eq:multiplicity}.
\end{eqnarray}
Here, the quantities $p'_i$ denote baryon momenta in the individual BB
center-of-mass system
which have to be transformed into the laboratory frame or the midrapidity
frame, respectively, and $s = ( p_1 + p_2)^2$ is the squared invariant energy
of
the collision. Finally, the antiproton invariant differential cross section
is obtained by integrating the differential multiplicity
(\ref{eq:multiplicity})
over the impact parameter b.

In order to proceed in the evaluation of (\ref{eq:multiplicity}) we assume,
as in refs. \cite{shor,danielewicz,Batko}, that the differential
elementary antiproton
cross section is proportional to the phase-space available for the final state:
{\samepage
\begin{eqnarray}
\lefteqn{
   E_3 \, E_4 \, E_5 \, E_{\bar{p}} \frac{d^{12}
   \sigma_{BB \rightarrow \bar{p}}(\sqrt{s})}
    {d^3p_{\bar{p}} d^3p_3 d^3p_4 d^3p_5}  }     \nonumber  \\
& & = \sigma_{BB \rightarrow \bar{p}}(\sqrt{s})
    \frac{1}{16 \, R_4( \sqrt{s} ) }
    \delta^4(p_1 + p_2 - p_3 - p_4 - p_5 - p_{\bar{p}} )
                                                  \label{eq:differential}.
\end{eqnarray}    }

The factor $1/R_4(\sqrt{s})$ is the 4-body phase-space integral
\cite{byckling} and has been included in order to normalize the
differential distribution. It should be noted that the  4-body
phase-space integral contrary to ref. \cite{Batko} now strongly
depends on the quasi-particle properties in the medium defined by eqs. (2) -
(5) and (10). The total cross section for antiproton production
$\sigma_{BB \rightarrow \bar{p}}(\sqrt{s})$ has been extracted from
the experimental data corresponding to the inclusive process $pp \rightarrow
\bar{p} +  X$. Unfortunately, there are no data
available at $\sqrt{s}-4m < 1$ GeV so that we perform an
extrapolation to lower energies. In line with \cite{Batko,Cass92,Faessl}
 we adopt the  parametrization:
\begin{equation}
\sigma_{pp \rightarrow \bar{p}+X}(\sqrt{s}) =
0.01 \: (\sqrt{s} - \sqrt{s_0} )^{1.846} \; [mb]       \label{eq:total}
\end{equation}
and assume that $\sigma_{pp \rightarrow \bar{p}+X}(\sqrt{s}) =
\sigma_{BB \rightarrow \bar{p}p+NN}(\sqrt{s})$. Whereas $\sqrt{s_0} = 3.7532
\ GeV$ for free particles, we replace $\sqrt{s_0}$ by the corresponding
threshold $\sqrt{s_0(\rho,p_i)}$ for the quasi-particles as defined by
eqs. (2) - (5) in the medium. Though the latter assumption is not
controlled by experimental data so far - and can hardly be measured - it is
well in line with the concept of a phase-space dominated elementary
production cross section.

Apart from the perturbative production scheme discussed above, the
antiprotons produced in individual baryon-baryon collisions can be
annihilated on their way out of the dense nuclear medium into the
continuum. Since a proper treatment of $\bar{p}$-annihilation is
decisive for a comparison with experimental data, we perform this task
nonperturbatively in the following way: For each baryon-baryon collision
event $i$ at space-time position $x_i$ we evaluate the differential
production probability $P_i(\bar{p}_j)$  for
an antiproton with momentum $\bar{p}_j$ on a grid in momentum space.
 Then each grid point in momentum
space $j$ (for each collision event $i$) is represented by an antiproton
testparticle which is propagated in time under the influence of its
selfenergies (10) according to the respective equations of motions
for point particles which provides individual trajectories $r_{ij}(t)$.
In order to account for annihilation, the individual probabilities
$P_i(\bar{p}_j;t)$ are integrated in time according to the following
rate equation
\begin{equation}
{d \over dt} P_i(\bar{p}_j;t) = - {4 \over {(2 \pi)^3}} \int d^3 p \
v_{12} \sigma_{ann}(\bar{p}_j, p)
 \ f(r_{ij}(t), p;t) P_i(\bar{p}_j;t)   \label{eq:ann}
\end{equation}
where $v_{12}$ is the Lorentz-invariant relative velocity, $f(r,p;t)$ the
baryon phase-space distribution from eq. (1) and $\sigma_{ann}$ the
$\bar{p}$ annihilation cross section which is taken from the experimental
data \cite{annihilationdata} and parametrized as a function
 of $\sqrt{s} - \sqrt{s_0}$.
Though it will remain a matter of debate, how $\sigma_{ann}$ might change
in the nuclear medium, we follow our concepts above and use the free
annihilation cross section, however, modify $\sqrt{s_0}$ in the medium
according to the quasi-particle properties described above. We note that the
validity of this concept can in part be controlled via the experimental
mass dependence of the $\bar{p}$ yield when comparing light and heavy
systems.  For further details we refer the reader to ref. \cite{Teis}.

We have applied the above mentioned formalism to evaluate the antiproton
cross section for the reactions p + $^{12}C$ and p + $^{63}$Cu at
bombarding energies of 5, 4,  and 3.5 GeV. The
corresponding invariant cross sections in comparison with the data of
ref.  \cite{KEK} are shown in Fig. 2 (a + b) as a function of
the momentum of the emitted antiproton in the lab frame at $\Theta = 0^o$,
assuming free antiprotons, i.e. $g_s^{\bar{p}}, g_v^{\bar{p}} = 0$.
The calculations slightly underestimate the experimental data, but
already approximately reproduce the shape of the momentum-spectra as well as
the dependence on bombarding energy and mass. When adopting a slightly
attractive scalar $\bar{p}$ selfenergy of - 50 to - 100 MeV the reproduction
of the data improves at all energies significantly which is exemplified for
4.0 GeV by the dashed line in Fig. 2. We note that in the above comparison
we cannot distinguish between scalar and vector antiproton selfenergies because
both yield similar results for the $\bar{p}$ spectrum if the same
Schr\"odinger-equivalent optical potential is achieved.
Furthermore, when using antiproton selfenergies in line with the relativistic
mean-field theories \cite{Serot}, i.e. changing only the sign of
 nucleon vector potential, we overestimate the $\bar{p}$ yield by more than
an order of magnitude at all energies for both systems.

We now turn to the nucleus-nucleus case. The calculated antiproton invariant
differential cross section for the reaction $^{28}$Si+$^{28}$Si at 2.1 GeV/A
and Ni + Ni  at 1.85 GeV/u is shown in Figure 3 in comparison
to the experimental data of ref.\cite{BEVALAC1} and ref. \cite{GSI};
the upper lines represent the results of  the calculations for
free antiprotons without including any reabsorption. When taking care of
antiproton annihilation according to eq. (14) the yields drop to the lower
full lines which now underestimate the data sizeably. However, using
attractive scalar (or vector) selfenergies at $\rho = \rho_0$ of
about - 100 to -150 MeV we nicely reproduce the data again.
 Since the two systems studied differ quite substantially
in mass we infer that the description of $\bar{p}$ annihilation appears to be
sufficiently accurate.

We note in passing that the contribution due to collisions involving resonances
is a factor 10 larger than those involving only nucleons
thus confirming again the main results of ref. \cite{koch,Batko}. A detailed
investigation of the baryonic decomposition for the present reactions is
given in ref. \cite{Teis}.

 The different
value for the attractive antiproton field at $\rho = \rho_0$
in p + A and A + A reactions is
due to the fact that in p + A collisions the antiprotons move with
momenta of 1 - 2 GeV/c with respect to the nuclear medium, whereas in
A + A collisions the antiprotons are almost at rest in the nucleus-nucleus
center-of-mass frame. In view of uncertainties of our present studies with
respect to the elementary production
cross sections close to the thresholds we provide
areas for the antiproton Schr\"odinger equivalent potential at $\rho = \rho_0$
in Fig. 4, as extracted from the comparison with the experimental
data for p + A and A + A reactions. These areas are far from the values
expected by charge conjugation from the familiar $\sigma - \omega$ model
\cite{Serot} (dashed line) and thus exclude
 relativistic mean-field models with the
same parameter-sets for nucleons and antinucleons. However, our extrapolated
values are well in line with a Schr\"odinger-equivalent potential (solid line
in Fig. 4) as extracted from the dispersion relation (P: principle value)
\begin{equation}
Re(U^{\bar{p}}(E,\rho_0)) =  {1 \over \pi} P  \int dE'
{Im(U^{\bar{p}}(E',\rho_0))   \over E - E'}
\label{disp}
\end{equation}
whereas the imaginary part is determined from the annihilation rate at
density $\rho_0$ according to
\begin{equation}
2 Im(U^{\bar{p}}(E,\rho_0)) = - {p \over E} \ \sigma_{ann}(E) \  \rho_0
\label{rate}
\end{equation}
with $E = \sqrt{p^2+m^2}$ and $\sigma_{ann}$ from
 ref. \cite{annihilationdata}. The real part of the $\bar{p}$-potential
(as shown in Fig. 4 by the solid line) is well in line with the
 potential analysis for $\bar{p}$ + A reactions \cite{Dover}.

In summary, we have evaluated the differential cross section for $\bar{p}$
production for p-nucleus and nucleus-nucleus reactions in the subthreshold
regime by considering incoherent on-shell baryon-baryon production processes
involving nucleons and  $\Delta$'s  with their in-medium quasi-particle
properties and treating $\bar{p}$ annhilation nonperturbatively. The
quasi-particle properties of the nucleons are fixed within our study by
the nuclear saturation properties, the proton-nucleus empirical potential
as well as Dirac-Brueckner calculations at higher density. In comparing our
calculations to the most recent data from KEK \cite{KEK} and  SIS
\cite{GSI} we find a consistent description of the experimental results
with a rather weak attractive potential for the antiprotons which is almost
perfectly in line with a dispersive potential extracted from the
dominant imaginary part of the antiproton selfenergy due to annihilation.
Though we cannot exclude  uncertainties due to uncertain
 elementary production cross sections we can infer, that relativistic
mean-field theories - which predict antiproton selfenergies according to
charge conjugation (and violate unitary) - are definitely inadequate
for the description of antimatter in a dense baryonic environment.

We note in closing that the antiproton production studies at the AGS
\cite{AGS1,AGS2,AGS3} around 15 GeV/u might
yield further information on the dynamics and selfenergies of antiprotons
at even higher densities although these processes occur far above threshold.

\vspace{1cm}
\noindent
The authors acknowlegde valuable discussions with A. Gillitzer, P. Kienle,
W. K\"onig and A. Schr\"oter as well as their information on experimental
 results prior to publication.

\section*{Figure Captions}
\newcounter{figno}
\begin{list}%
{\underline{Fig.\arabic{figno}}:}{\usecounter{figno}
         \setlength{\rightmargin}{\leftmargin}}

\item    Scalar and vector selfenergies $U_s(p)$ and $U_0(p)$ for nucleons
         at different densities in units of $\rho_0 \approx 0.17 fm^{-3}$
         as used in the relativistic transport equation (1).

\item    Invariant cross section for antiproton production in the reactions
         p+$^{12}C$ and
         p+$^{63}$Cu at $\Theta=0^o$ as a function of the antiproton momentum
         p in the
         lab system. The experimental data are taken from
         ref.\ \protect\cite{KEK} and correspond to bombarding energies
         of 5.0 GeV , 4.0 GeV  and 3.5 GeV. The full lines represent
         calculations for free antiprotons.
         The dashed lines indicate the result for an antiproton selfenergy of
         - 100 MeV at 4.0 GeV.

\item    Invariant cross section for antiproton production in the reaction
         $^{28}$Si+$^{28}$Si at 2.1 GeV/u and Ni + Ni at 1.85 GeV/u for
         $\Theta=0^o$ as a function of the momentum p of the
         emitted antiproton in the lab-system. The experimental data have been
         taken from
         refs.\protect\cite{BEVALAC1} and \protect\cite{GSI}, respectively.
          The upper lines indicate the calculated cross section for free
         antiprotons without reabsorption whereas the lower solid line is
         obtained when including $\bar{p}$ annihilation. The dashed line
         represents the cross section adopting an attractive potential
         of the antiproton of - 150 MeV.

\item    Comparison of our extracted values for the Schr\"odinger
         equivalent antiproton potential from p + A and A + A reactions
         with the prediction from the $\sigma - \omega$ model (dashed line)
         and the dispersive potential according to eq. (15) (solid line).

\end{list}

\end{document}